\documentclass[fleqn,10pt]{SelfArx} 

\usepackage[english]{babel} 
\usepackage{epsfig}
\usepackage{lipsum} 

\def\apj{Astrophys. J.}
\def\apjl{Astrophys. J. Lett.}
\def\apjs{Astrophys. J. Suppl.}
\def\mnras{Mon. Not. R. Astron. Soc.}
\def\araa{Annu. Rev. Astron. Astrophys.}
\def\aap{Astron. Astrophys.}
\def\aj{Astron. J.}
\def\nat{Nature}
\def\nar{New Astron. Rev.}

\definecolor{color1}{rgb}{0.0,0.0,.0} 
\definecolor{color2}{rgb}{0.925, 0.956, 0.992} 
\definecolor{color3}{rgb}{1.0,1.0,1.0} 
\definecolor{color4}{rgb}{0.0,0.0,0.0} 
\definecolor{colorbox}{rgb}{0.925, 0.956, 0.992}
\definecolor{colorheader}{rgb}{0.33,0.41,0.47} 

\usepackage{hyperref} 
\hypersetup{hidelinks,colorlinks,breaklinks=true,urlcolor=color2,citecolor=color1,linkcolor=color1,bookmarksopen=false,pdftitle={Title},pdfauthor={Author}}

\JournalInfo{Invited Review for Nature Astronomy} 
\Archive{} 

\PaperTitle{Impact of supermassive black hole growth on star formation} 

\Authors{C. M. Harrison\textsuperscript{1,2}*} 
\affiliation{\textsuperscript{1}\textit{European Southern Observatory, Karl-Schwarzschild-Str. 2, 85748
   Garching b. M{\"u}nchen, Germany}} 
\affiliation{\textsuperscript{2}\textit{Centre for Extragalactic Astronomy, Durham University, South
Road, Durham, DH1 3LE, U.K.}} 
\affiliation{*ESO Fellow; c.m.harrison@mail.com} 


\Abstract{Supermassive black holes are found at the centre of massive
galaxies. During the growth of these black holes they light up to become visible as active galactic nuclei (AGN) and release extraordinary amounts of energy across the electromagnetic spectrum. This energy is widely believed to regulate the rate of star formation in the black holes' host galaxies via so-called ``AGN feedback''. However, the details of how and when this occurs remains uncertain from both an observational and theoretical perspective. I review some of the observational results and discuss possible observational signatures of the impact of super-massive black hole growth on star formation.}

\begin{document}
\fontfamily{lmss}\selectfont
\flushbottom 

\maketitle 


\thispagestyle{empty} 


\section*{Introduction} 

\addcontentsline{toc}{section}{Introduction} 

The discovery that all massive galaxies host a central super-massive
black hole rates among the most momentous in modern astronomy. These
black holes, with masses ranging from hundreds of thousands to
billions of times that of our Sun
($\approx$10$^{5}$--10$^{10}$\,M$_{\odot}$), primarily grow through
periods of radiatively-efficient accretion of gas when they consequently become visible as
AGN\cite{Soltan82,Marconi04}. Historically AGN were considered rare but fascinating objects to study in their own right, yet over the
last two decades these phenomena have moved to the fore-front of galaxy
evolution research. This is partly due to a number of remarkable
observations that show that black hole masses are tightly correlated
with host-galaxy properties, despite a difference of several orders of
magnitude in physical size scales\cite{Kormendy13}. However, arguably the most influential factor
in the explosion of interest in AGN are the results from galaxy
evolution models. 

Most galaxy formation models require AGN to inject energy or momentum
into the surrounding gas (see Box~1) in the most massive galaxies
(i.e., with stellar masses $M_{\rm stellar}\gtrsim10^{10}$\,M$_{\odot}$) in order to reproduce many key observables of galaxy populations and
intergalactic
material\cite{Valageas99,Croton06,Somerville08,Ciotti10,Gaspari11,Dubois13,Vogelsberger14,Crain15,King15}
(Fig.~1). These observables include: the ``steep'' relationship between X-ray luminosity and X-ray temperature observed for the gas in the
intra-cluster medium within groups and clusters\cite{Markevitch98}; the ``low'' rate
of gas cooling in galaxy clusters\cite{Fabian94}; the inefficiency of star
formation in the most massive galaxy haloes\cite{Behroozi13} 
(Fig.~1); the tight relationships between black hole
masses and galaxy bulge properties\cite{Kormendy13} and the
formation of quiescent bulge-dominated massive ``red'' galaxies that
are no longer forming stars at significant levels\cite{Strateva01}.

\begin{figure}
\centerline{\psfig{figure=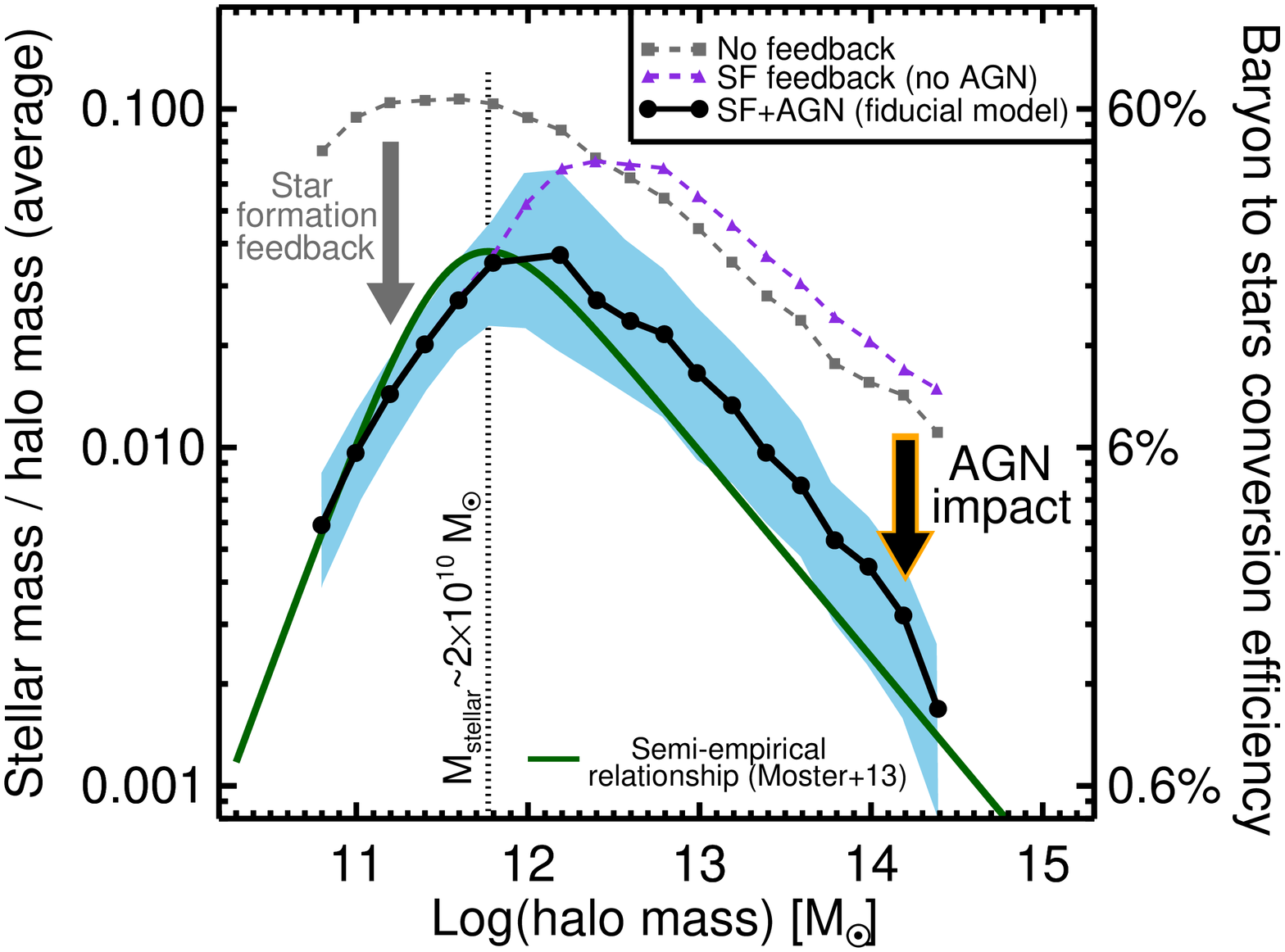,width=0.35\textwidth,angle=90}}
\fontfamily{phv}\selectfont{\small Figure~1 | The ratio of stellar mass to halo mass as a function of halo mass for three different runs of a
  simulation\protect\cite{Somerville08} and for the semi-empirical
  relationship\protect\cite{Moster13}.} {\footnotesize The shaded region shows the 16th and
84th percentiles of the fiducial model that includes energy injection from
AGN and star formation (SF). The right y-axis shows the efficiency
for turning baryons into stars ($M_{\rm stellar} /[f_{b}*M_{\rm
  halo}]$; where the factor of $f_{b}=0.17$ is the cosmological baryon
fraction). The impact of including star formation feedback in the
model is to reduce the efficiency of converting baryons into stars in
low mass haloes. For massive haloes, energy injection
from AGN is required in order to reduce these efficiencies. Such
effects are required in most models in order to reproduce many
observable properties of the
massive galaxy population.}
\end{figure}

\begin{figure*}
\noindent\fcolorbox{white}{colorbox}{%
\begin{minipage}{0.5\textwidth}
\centerline{\psfig{figure=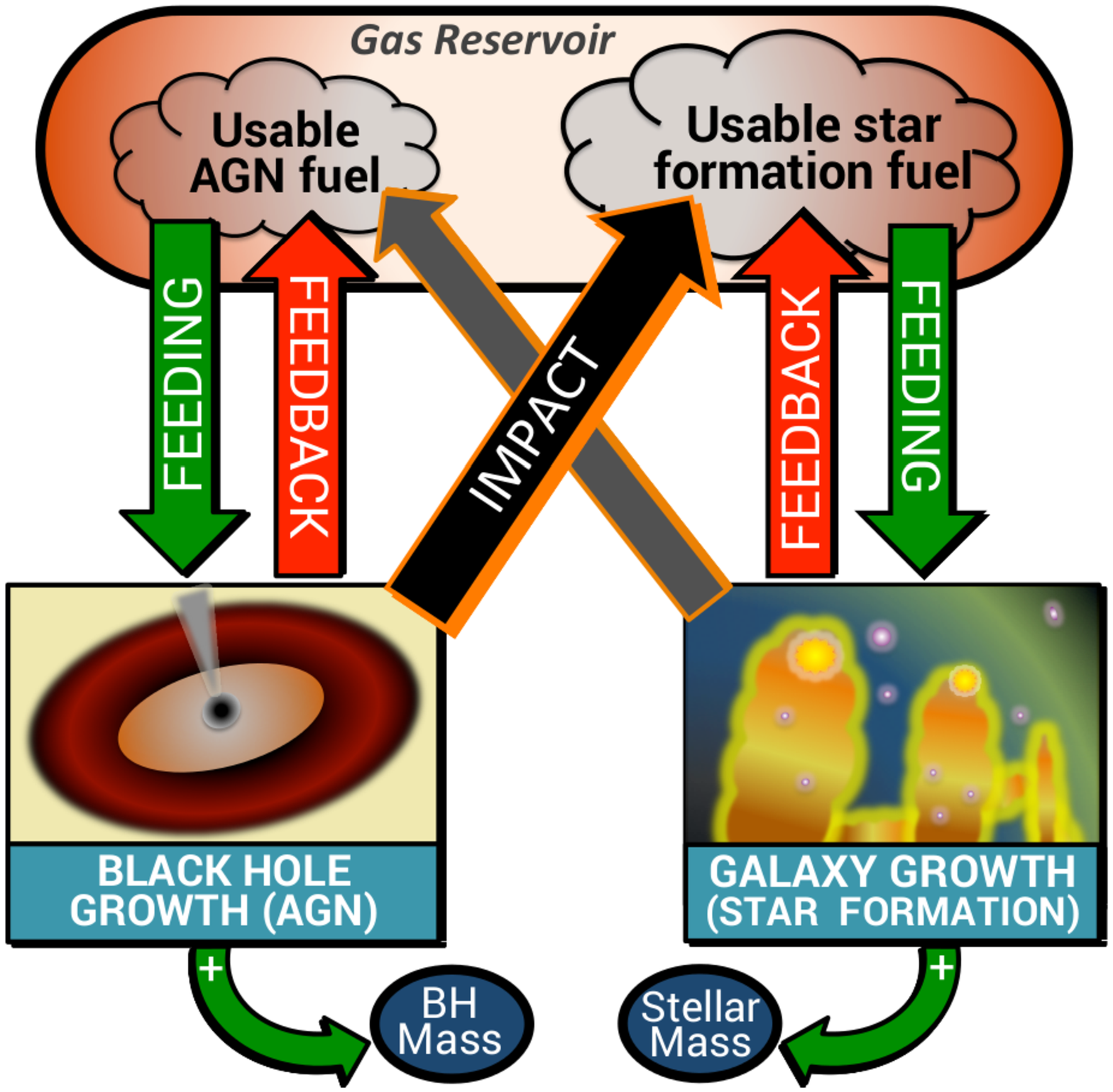,width=\textwidth,angle=0}}
\end{minipage}
\begin{minipage}{0.5\textwidth}
\fontfamily{phv}\selectfont{\small Box~1 | A schematic diagram to illustrate the relationships
    between fuel supply, galaxy growth and black hole
    growth.} 

\vspace{0.1cm}\noindent {\footnotesize Both AGN and star formation are fuelled by cold gas that originates from a shared
  (potentially hot) {\em gas reservoir} inside the galaxy halo. This
  gas reservoir can be fed by
  gas-rich mergers, by recycled material from internal galactic
  processes and by accretion of gas from intergalactic material. The amount of gas and the ability for this gas to cool determines the
  amount of usable fuel that can be used for {\em feeding} black hole
  growth and star formation. In the case of providing the fuel for black hole growth the
  material has the additional challenge of losing sufficient angular momentum to reach the inner
  sub-parsec region of the galaxy. Both processes are known to inject energy and momentum (via
  radiation, winds and jets) that can reduce the availability of
  usable fuel through ionising, heating, shocking or expelling material, and hence provide self-regulatory
  {\em feedback} mechanisms. A key component of most
  galaxy formation models is that these two processes can also have a
  positive or negative {\em impact} on the usable fuel supply for the other process (black and grey arrows). The focus of
  this article is observational results on the
  impact of black hole growth on star formation.
}
\end{minipage}
}%
\end{figure*}

AGN are an attractive solution in models to supply the energy required to
explain the observations. By releasing $\approx$10\% of the rest-mass energy of accreted material, they are phenomenal energy
sources\cite{Shapiro83,Marconi04}. For example, during the formation of a $\approx$10$^{8}$\,M$_{\odot}$ black
hole $\approx$10$^{54}$\,Joules of energy is released, which is
two-to-three orders of magnitude more energy than the binding energy
of a typical host galactic bulge and is
comparable to the thermal energy of the gas in the
galaxy halo. Consequently, if only a small fraction
of this energy is able to couple to the gas it will be capable of
regulating black hole growth and the star formation in the host galaxy
(see Box~1).

Whilst it is theoretically attractive to invoke AGN as a mechanism to
regulate the rate of star formation in massive galaxies,
this can only be credible if backed up by observational evidence. The observational
task is to assess if and how accretion energy couples to gas and what
resulting impact this then has on star formation in the AGN host galaxies.

\section*{Methods of energy injection by AGN}
The energy released by black hole accretion (AGN) may be radiative
(i.e., energetic photons) or mechanical (i.e., energetic
particles)\cite{Cattaneo09,Ciotti10,Weinberger17}. In models, radiative energy injection is
sometimes called ``quasar'' or ``wind'' mode and is usually associated
with high Eddington ratios ($\gtrsim$0.01; i.e., mass accretion rates
that are $\gtrsim$1\% of the theoretical maximum ``Eddington
limit''). In contrast mechanical energy injection is sometimes called
``radio'' or ``jet'' mode and is associated with low Eddington
ratios. Early analytical models invoked galaxy-wide gas outflows, initially
launched by accretion radiation coupling to the gas on small scales, to explain the
observed scaling relationships between galaxies and black
holes\cite{Silk98,King03}. In hydrodynamical simulations energy
injection from AGN is often crudely implemented; for example, by assuming
a small fraction of the total radiative luminosity of
accreting black holes couples thermally to the surrounding gas, with the result of expelling
material from the host galaxy in an outflow and suppressing star
formation\cite{Springel05,Hopkins06}. However, recently simulations have
incorporated more complex prescriptions for ``feeback'' by invoking
and testing multiple modes of energy injection\cite{Ciotti10,Choi15,Weinberger17}. Observational
constraints on the different feedback prescriptions are a critical
test of these models. 

Based on the above, it is convenient to classify observed AGN into two broad categories: those for which their
energetic output is predominantly radiative (radiative AGN) and those
for which it is predominantly mechanical (mechanically-dominated
AGN)\cite{Best12}. Radiative AGN are luminous in X-rays, optical and/or infrared
emission (sometimes also in radio emission) and are rare among the
galaxy population as a whole ($\lesssim$ a few percent)\cite{Aird12}. Mechanically-dominated AGN are
usually identified through luminous radio emission\cite{Heckman14};
however, those identified are found in the most massive systems and a
rare subset of all galaxies which host low black hole accretion rates\cite{Best12}.

Mechanically-dominated AGN are pre-dominantly found in the most
massive galaxies (\allowbreak{$M_{\rm
    stellar}\gtrsim$10$^{11}$\,M$_{\odot}$}) with old stellar
populations, at least in the local Universe, whilst radiative AGN are
most common in galaxies with on-going star-formation and younger stellar populations at
all cosmic epochs\cite{Hickox09,Heckman14,HernanCaballero14}. Consequently, these two
categories of AGN may represent distinct evolutionary phases and/or
distinct black hole accretion mechanisms depending on the host galaxy mass and
environment\cite{Tasse08,Best12}. Therefore, when assessing the impact
of AGN on star formation it is important to consider
these AGN types separately. Care is especially required for AGN that are
identified through luminous radio emission that are increasingly
more mechanically dominated towards later cosmic times (i.e.,
redshifts $z\lesssim1$) and are increasingly
more radiatively dominated at early cosmic times\cite{Best14,Padovani15}.

Although the details remain uncertain there is compelling observational
evidence, at least in the local Universe and in the densest
environments, that radio jets driven by mechanically-dominated AGN can
maintain host galaxy star formation at low levels. This
is achieved by suppressing the ability for hot gas to cool (see Box~1) and has
been reviewed extensively in the
literature\cite{Cattaneo09,McNamara12,Fabian12}. However, it is not
yet fully understood what role AGN play in less dense
environments\cite{Cattaneo09,Donoso10} or if gas needs to be ejected during
earlier AGN episodes for these mechanically-dominated AGN to be
effective at regulating gas cooling\cite{McCarthy11}. Furthermore, for these massive galaxies most of the galaxy and black hole
growth occurred at earlier cosmic epochs than where this radio jet
heating has been identified\cite{Heckman04,Thomas05}
and it is not yet clear what quenched the earlier high rates of star
formation in these systems\cite{Schawinski14}. 

To work towards addressing the outstanding issues raised above and to fully characterise the
impact of AGN on star formation it is crucial to study and understand
the role of {\em radiative} AGN. This is particularly true at early cosmic times (i.e., $z\gtrsim0.5$), when significant levels of black
hole and galaxy growth were occurring. The remainder of this review
will focus on the observational evidence for the impact
of radiative AGN on star formation. As described in Box~2 a common theme throughout the
following sections will be awareness of the relative and uncertain
timescales of: (1) visible AGN episodes; (2) star formation episodes
and; (3) the impact of AGN energy injection on star formation.

\begin{figure}
\noindent\fcolorbox{white}{colorbox}{%
\begin{minipage}{0.5\textwidth}
\noindent\fcolorbox{colorbox}{white}{%
\begin{minipage}{0.973\textwidth}
\centerline{
  \psfig{figure=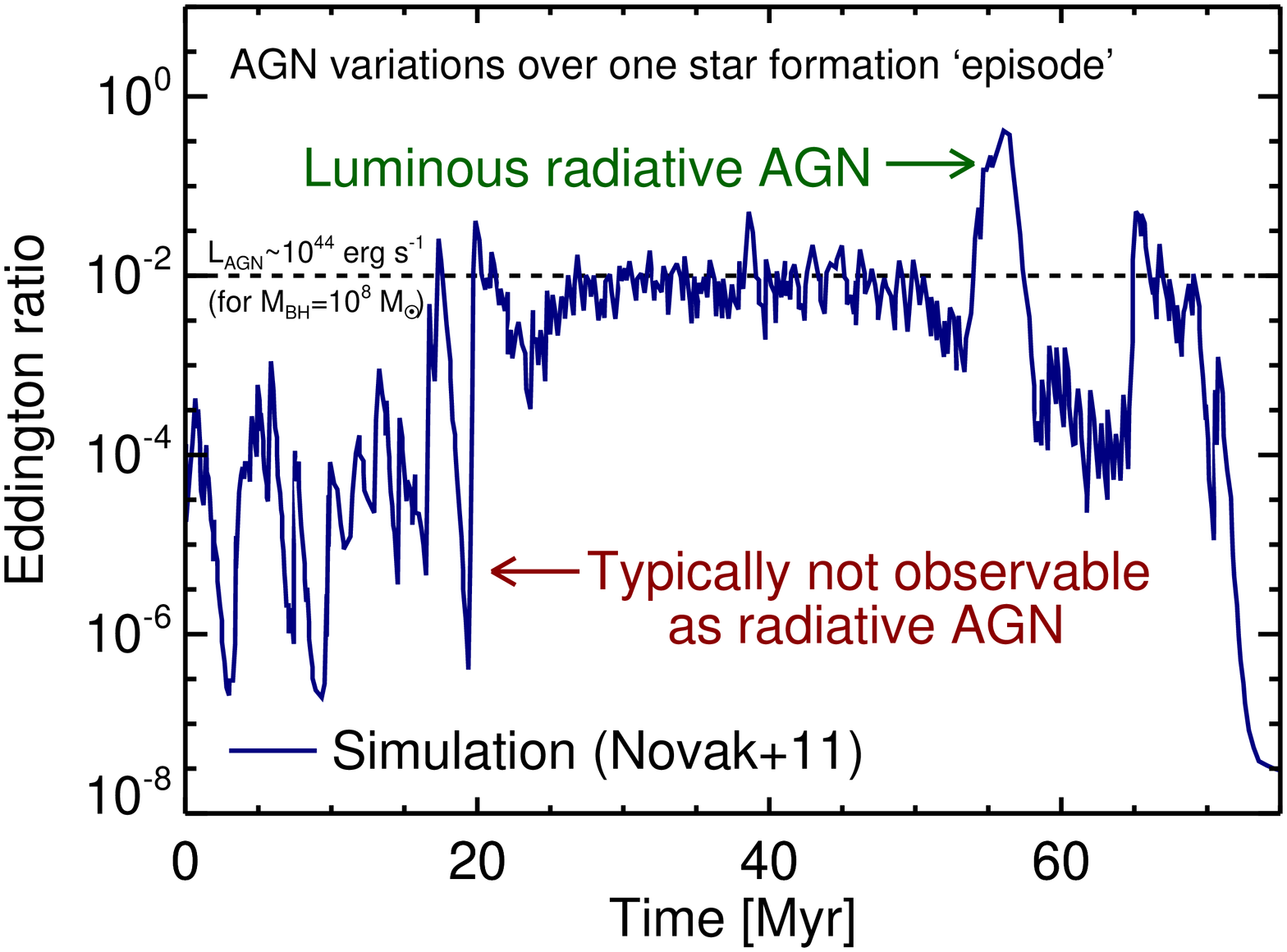,width=0.75\textwidth,angle=90}
}
\end{minipage}
}
\vspace{0.05cm}

\fontfamily{phv}\selectfont{\small  Box~2 | Eddington ratio versus times for an example simulation
    of an AGN to illustrate variability.} 

\vspace{0.05cm}\noindent{\footnotesize As discussed in detail in \protect\cite{Hickox14} various
  observational work indicates that AGN luminosities ($L_{\rm AGN}$), in particular
  those derived from optical and/or X-ray continuum measurements that trace
effectively {\em instantaneous} mass accretion rates, vary on orders of magnitude on times scales much
shorter than the typical timescale of star formation episodes
($\gtrsim$100\,Myrs). Similar results are reached by AGN simulations;
for example, the figure presents the results of the Eddington
    Ratio (proportional to the mass accretion rate and AGN luminosity) as a function of
    time for an example hydrodynamical simulation\protect\cite{Novak11}. This model predicts that accretion
  rates can vary by several orders of magnitude on timescales of
  $\lesssim$1\,Myr. Consequently, measured AGN luminosities may provide little information on the cumulative
energy released over the relevant timescales for star
formation. Understanding the timescales traced by the various AGN
luminosity indicators is crucial for our interpretation of the impact
of AGN determined from observations. Furthermore, the relative timescales of a visible luminous AGN and the time taken for any resulting impact on the
observed star-formation rates are very uncertain. Crucially, even when the AGN is responsible for
enhancing or decreasing the star-formation rate in the host galaxy, it is most likely
that the AGN luminosity will vary much more rapidly than the
star-formation rates\cite{Zubovas13,Thacker14}. Such
effects are important to consider when assessing the impact of AGN on
star formation through observations.}
\label{fig:variab} 
\end{minipage}
}
\end{figure}

\section*{Observing the mechanism of energy injection by radiative
  AGN}
A common approach towards understanding the impact of AGN on star formation is
to search for and to characterise a mechanism by which AGN are injecting
energy and/or momentum into the gas in their host galaxies (see
Box~1). For example, outflows may remove gas from the host galaxy and
have the effect of suppressing star formation. Alternatively, AGN might kinematically disturb,
compress, shock and/or heat the gas via outflows or jets and
consequently reduce or enhance the ability for the gas to form
stars. It is not the purpose of this review to comprehensively cover the huge amount of observational work on outflows or jets driven by radiative AGN (see
\cite{Veilleux05,Alexander12,Fabian12,King15}). However, below I focus on some of the
observational work that specifically investigates the impact that these outflows
may have on star formation.

Radiatively-driven AGN outflows are known to be common on small spatial scales,
i.e., close to the accretion disk, in the form of the extremely high speed winds that are identified in X-ray and ultra-violet spectroscopy
(up to $v\approx0.1$--$0.2\times$ the speed of
light\cite{Ganguly08,Tombesi10}). These winds have the
potential to provide the feedback mechanism for self-regulating black hole growth
(Box~1). Furthermore, lower velocity outflows in multiple gas phases (i.e.,
outflows of ionised, neutral and molecular gas) have been identified using
one-dimensional spectra of AGN host galaxies and are more likely to
be associated with host galaxy gas\cite{Rupke05,Dunn10,Sturm11,Mullaney13}. In some cases these
outflows are inferred to be located on 100s--1000s of parsec scales by
applying a variety of modelling techniques, such as
radiative transfer and photoionization models, to the information
extracted from the spectra\cite{Sturm11,Dunn10}. What is even more pertinent is the direct
detection of outflows on kiloparsec scales, in multiple gas phases, using spatially-resolved kinematic
measurements\cite{Veilleux05,Maiolino12,Cicone14,Harrison14,Nesvadba16}. Only if AGN can influence gas on
$\gtrsim$kiloparsec scales will they be able to have a significant
impact upon the galaxy-wide star formation in their host
galaxies. Understanding how AGN accretion disk winds couple to multi-phase gas on galaxy-wide scales
is an on-going observational and theoretical challenge\cite{Tombesi15,Feruglio15,King15}.

\begin{figure}
\centerline{\psfig{figure=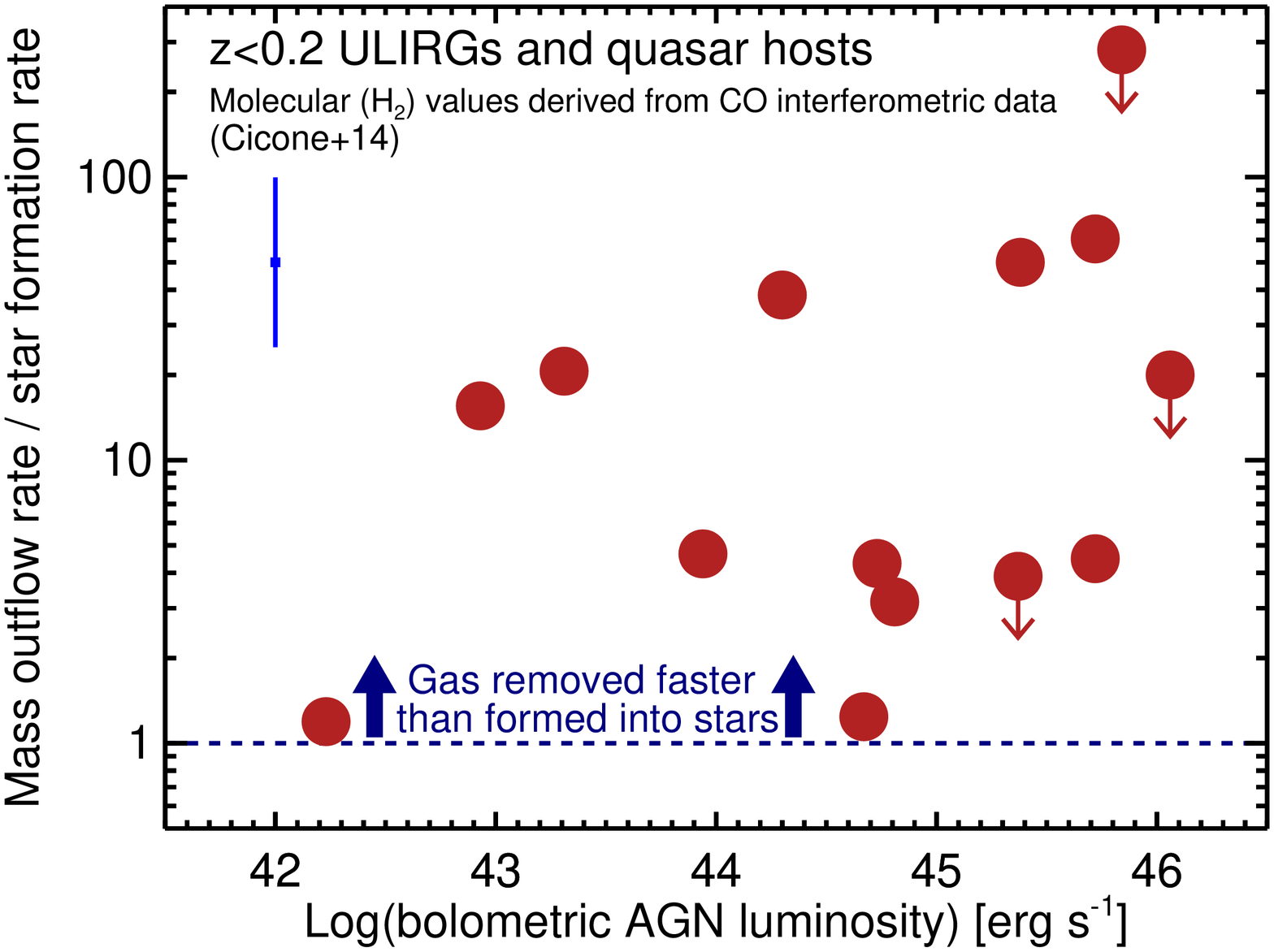,width=0.38\textwidth,angle=90}
}
\fontfamily{phv}\selectfont{\small Figure~2 | Ratio of H$_{2}$ mass outflow rate to star formation rate as a
    function of AGN luminosity for low redshift ULIRGs and quasar host
  galaxies\protect\cite{Cicone14}.} {\footnotesize These measurements
imply that molecular gas is being removed by AGN-driven outflows faster than it can be formed into stars. A representative error bar
is shown in the top left, but this does not include the large and
unknown uncertainty on converting CO to H$_{2}$ masses\protect\cite{Cicone14}.}
\end{figure}

Example evidence that AGN-driven outflows may have a
significant impact upon star formation is that the measured mass outflow rates
of molecular outflows in rare low redshift ultra-luminous infrared galaxies (ULIRGs) and quasar host galaxies appear to exceed the concurrent
star-formation rates\cite{Cicone14} (see Fig.~2). Consequently star-forming material appears to be
being removed more rapidly than it can be formed into stars in these galaxies. Similar arguments
have also been made for more typical AGN host galaxies using a variety
of gas tracers\cite{Liu13,Harrison14}. However, there are various difficulties involved
with deriving the measurements and performing these analyses, with
dramatically different results possible when applying different
techniques, making different assumptions or when using different gas
tracers\cite{Harrison14,Husemann16,Rupke13,GonzalezAlfonso17}. Furthermore, understanding the
timescale on which these outflows occur is troublesome and crucial for
the interpretation on the long term impact of these outflows\cite{GonzalezAlfonso17}. Particularly challenging is making
these measurements beyond the local Universe, where, without excellent observations using adaptive optics or
interferometers the spatial resolution can be comparable to, or higher
than, the spatial extents of the outflows.

Towards a more direct indication that AGN-driven outflows may influence
star formation, there have been observations of
a small number of distant luminous AGN ($z\approx1$--3) that show evidence
for an anti-correlation between the spatial location of an ionised outflow and the location of
narrow H$\alpha$ emission (a star-formation tracer)\cite{CanoDiaz12,Carniani16}. These results may indicate that star formation has been
reduced in the regions of the outflow, although an alternative
possibility is that these diffuse outflows preferentially escape away
from the dense star forming material\cite{Gabor14}. Indeed, AGN-driven
kiloparsec scale outflows are often found co-incident with high levels of on going star
formation\cite{Cicone14,Wylezalek15}. In some cases, observational papers have also reported evidence
of regions of enhanced star formation due to AGN-driven
outflows or jets, and even suppression
and enhancement working simultaneously in the same
galaxies\cite{Elbaz09,Cresci15}. 

Whilst much work has focussed on the idea that AGN should be able to
evacuate galaxies of star-forming material, studies of nearby galaxies
making use of (sub)-millimetre observatories have indicated that complete evacuation of cold molecular gas from a host galaxy is not a
pre-requisite to shut down an intense star formation episode. Systems with a large molecular gas reservoir can be
forming stars less efficiently than ``typical'' galaxies with the same
molecular gas mass, potentially due to the injection of turbulence
which inhibits the formation of gravitationally bound
structures\cite{Ho05,Guillard15,French15,Alatalo15}. In some sources
AGN seem to be the most likely energy source\cite{Alatalo15,Guillard15}. 

Observations have clearly identified that AGN can inject considerable
energy/momentum into their host galaxies and investigation into the
observable impact of this energy injection on star formation in
individual galaxies is on-going. However, one of the greatest on-going
challenges with these types of studies is to determine what long term impact AGN can have on their host galaxies. For example,
even if measured outflow rates are very high (e.g., Fig.~2) and/or the star formation efficiencies are very
low, it is not clear how long these episodes will last or if
re-accretion of material will trigger future star formation. Furthermore,
directly relating these episodes to the energy released by the central AGN is challenging due to the uncertain
timescales of visible AGN activity and the resulting measurable impact
(see Box~2). Insight may be obtained from {\em statistical} studies of
the star formation properties of galaxies with and without a visible AGN.

\section*{Star formation properties of radiative AGN host galaxies}
Towards assessing the impact of AGN on star formation, there has
recently been an abundance of studies investigating the star formation
rates of large samples of AGN host galaxies. Studies of purely
mechanically-dominated AGN, at least for the most radio luminous, consistently find that they reside in low star-formation rate host
galaxies\cite{Hardcastle13,Ellison16,Leslie16}. However, for radiative AGN the
conclusions have varied widely in the literature, with claims of star-formation rates that are: unrelated to AGN luminosity\cite{Mainieri11}, enhanced for the
most luminous AGN\cite{Lutz10}, inhibited for the most luminous AGN\cite{Page12} or both enhanced and reduced
depending on the wave-band used to trace the luminosity of the AGN\cite{Zinn13,Karouzos14}.

The conflicting conclusions for the star-formation rates of radiative
AGN can largely be attributed to the different samples and
approaches used. For example: (1) low numbers of the most
luminous AGN can lead to statistical fluctuations; (2) it is difficult to
convert photometric measurements into star formation rates (e.g., because of dust attenuation of optical and
ultra-violet emission and the challenges of removing the AGN contribution
to the emission at all wavelengths); (3) samples that
only consider AGN that are detected in far-infrared surveys will be biased towards higher
star-formation rates and (4) samples that are radio bright may contain
both high star-formation rate radiative AGN and low star-formation
rate mechanically dominated AGN. Another fundamental
factor to consider, is how the underlying correlations between star-formation rate and both redshift
and stellar mass are accounted for in each study. For example, a
positive correlation between star formation rate and AGN luminosity may
be driven by the fact that the most luminous AGN are hosted by the highest
stellar mass galaxies.

The studies that contain some of the largest samples of AGN host
galaxies, that have simultaneously taken into account redshift and
stellar mass and that have applied uniform techniques across their samples find
that average star-formation rates are independent of AGN
luminosity\cite{Rosario13b,Azadi15,Stanley15,Shimizu17} (Fig.~3). Does this result indicate that radiative AGN have no positive or negative impact
on galaxy-wide star formation rates? Addressing this question is non-trivial as it is extremely challenging to interpret the empirical
result. As described in Box~2 the relative timescale of an AGN to be
luminous compared to the timescale for any impact on the observed star
formation rates are very uncertain. Furthermore, some models suggest that AGN are unable to
have a direct impact upon {\em concurrent} star formation but instead the cumulative
effects of multiple AGN episodes may inhibit {\em future} star formation\cite{Gabor14}. With these aspects in mind, it clearly limits
what can be inferred from the star-formation rates of AGN without
complementary theoretical predictions.

\begin{figure}
\centerline{
  \psfig{figure=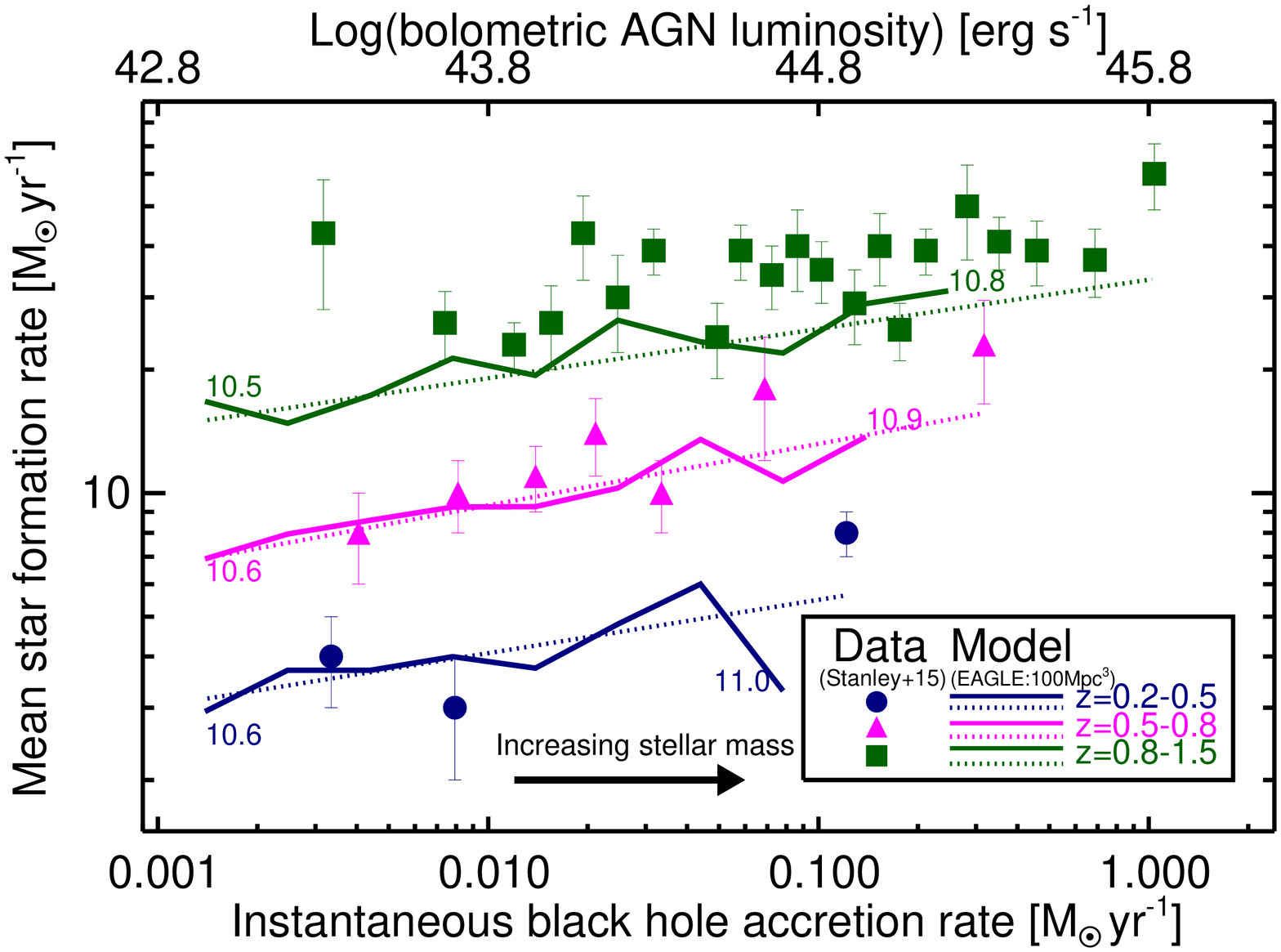,width=0.35\textwidth,angle=90}
}
\fontfamily{phv}\selectfont{\small  Figure~3 | Mean star formation rate versus instantaneous
  black hole accretion rate for a cosmological
  simulation\protect\cite{McAlpine17} and versus AGN luminosity (converted from X-ray luminosities) for
  observations\protect\cite{Stanley15}}. {\footnotesize The dotted lines are a linear fit to the
  running means for the model (solid curves). The logarithm of the
  average 30\,kpc aperture stellar masses (in
  stellar mass units) of the first and last bin are labelled; the
  slight increase in mean star-formation rate with increasing accretion rate is attributed to the
  increasing average stellar masses. Despite effective star-formation
  suppression by AGN in the model, this does not result in reduced
  average star-formation rates for the highest instantaneous black hole
  accretion rates (i.e., AGN luminosities).
}
\end{figure}

It is informative to obtain a prediction on the star formation rates of
AGN from a cosmological model that requires the suppression of star formation during periods of rapid
black hole growth to reproduce observable galaxy properties. For example, in agreement with
the data, the reference model of the EAGLE simulations (that includes
thermal energy injection from AGN)\cite{Schaye15} shows no evidence for
reduced average star formation rates with increasing black hole accretion
rate\cite{McAlpine17} as shown in Fig.~3. In Fig.~3: the star formation rates are
galaxy-wide, are averaged over 100\,Myrs to broadly match the
observed far-infrared measurements and are shifted up by 0.2\,dex,
to account for a systematic offset seen for all galaxies in the
simulation; the instantaneous black hole accretion rates are converted to
bolometric AGN luminosities assuming a radiative efficiency of 10\% (all details in \cite{McAlpine17}). Due to accretion rate variations
that happen more rapidly than the star formation rate variations, the
effects of star formation suppression does not result in a negative
trend in the star-formation rate versus AGN luminosity plane. Although based on a single model, this test
highlights that it is not possible to conclude a lack of impact by AGN
upon star formation based purely upon an empirical result where average star-formation rates are not reduced
for galaxies that host the most instantaneously luminous AGN. 

Further insight will be gained on this topic by analysing the full distributions of star
formation rates (not just simple averages) for radiative AGN host
galaxies\cite{Symeonidis13,Azadi15,Mullaney15,Leslie16,Ellison16}
in the context of theoretical predictions. Furthermore, further work using detailed spectra to assess
the star formation {\em histories} of AGN host galaxies, in tandem with specific model predictions on
how AGN and star formation interact, will also provide insight
into the observable signatures of the impact of AGN\cite{Smethurst16,Dugan14}. However, as I will suggest in the next section investigating the massive galaxy
population as a whole, irrespective of the presence of a luminous AGN, may yield some of the most informative results on the
impact of AGN on star formation.

\section*{Star-formation rates of massive galaxies}

As already described, it is a popular and effective method in galaxy formation models to invoke AGN to
reduce the star formation of the most massive
galaxies (Fig.~1). Even the most simple ``empirical'' galaxy formation models require some process to ``quench''
the most massive galaxies\cite{Peng10}. Therefore, insight into the
impact of AGN on star formation may be gained
from investigating the star-formation rates as a function of stellar
mass. In the star-formation rate versus stellar mass plane, galaxies are generally classed into two
categories; ``star-forming galaxies'' that follow a relatively tight positive relationship
between star-formation rate and stellar mass and ``quiescent
galaxies'' that fall below this relationship, where the fraction of quiescent
galaxies increases with stellar mass\cite{Strateva01,Brinchmann04,Whitaker14}.

Recent work has shown that star-forming galaxies with low stellar masses, i.e., below $\lesssim$ few
$\times$10$^{10}$\,M$_{\odot}$, follow an almost linear relationship
between average star-formation rate and stellar mass whilst more massive
star-forming galaxies, both with and without a luminous AGN, have a shallower
slope\cite{Whitaker14,Schreiber15,Cowley16} (Fig.~4).
This reveals that the star-formation rates per unit mass are smaller in the galaxies above this stellar mass
threshold. This effect is observed to already be in place $\approx$3\,Gyrs
after the Big Bang (redshift $z\approx2$) although the exact
form of the star-formation rate versus stellar mass relationship evolves with time\cite{Brinchmann04,Whitaker14,Schreiber15}
(Fig.~4). Consequently, it is a useful exercise to investigate the role of AGN in reducing the
relative growth rates of the most massive galaxies using model
predictions. 

\begin{figure}
\centerline{
  \psfig{figure=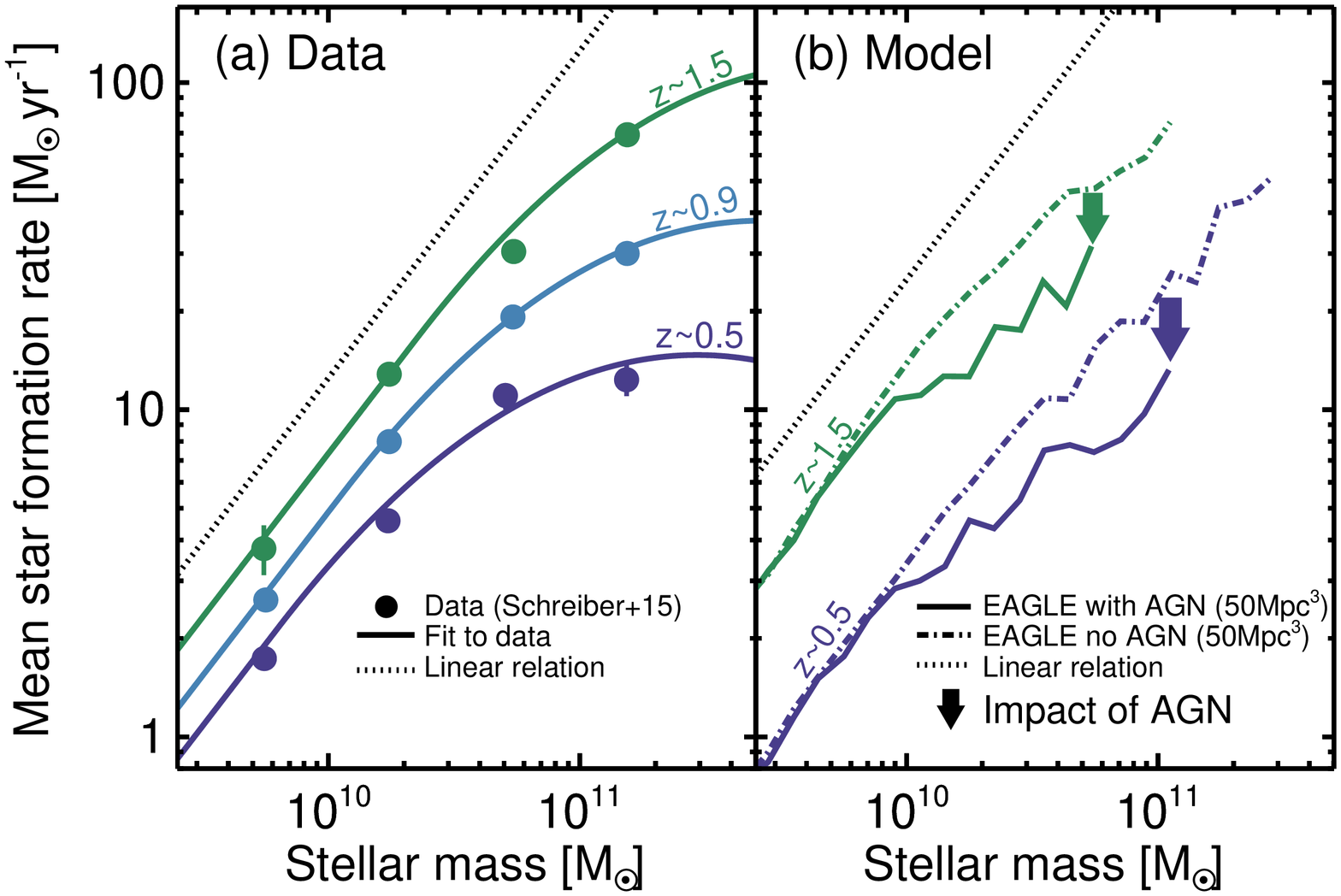,width=0.35\textwidth,angle=90}
}
\fontfamily{phv}\selectfont{\small Figure~4 | Mean star formation rate versus stellar mass for
    observed star-forming galaxies\cite{Schreiber15} (a) and
    galaxies in a cosmological model run both with and without
    AGN\protect\cite{Schaye15,Crain15} (b)}. {\footnotesize More
  massive galaxies form stars more rapidly; however, the highest-mass
  galaxies ($M_{\rm stellar}\gtrsim10^{10}$\,M$_{\odot}$) are observed to fall below a constant scaling
  relationship implying a reduction in the ability for the available baryons to be
  converted into stars\protect\cite{Whitaker14,Schreiber15}. In
  the model, the impact of AGN is to reduce star formation rates of
  high mass galaxies as well as to reduce the overall number of
  massive galaxies. Note that error bars are smaller than the data points in most cases\cite{Schreiber15}.
}
\end{figure}

Fig.~4 shows the running average star-formation
rate as a function of stellar mass of galaxies from two runs of the cosmological
hydrodynamical EAGLE simulations; the 50\,Mpc$^{3}$ box reference model (where AGN are
effective in regulating star formation) and an identical run, except
where AGN are ``turned off''\cite{Schaye15,Crain15}. Following
\cite{McAlpine17}, the star formation rates are total values and the
stellar masses are 30\,kpc aperture values (taken from the EAGLE
database\cite{McAlpine16}). Averages are only
calculated for stellar mass bins containing more than 15 galaxies. These two runs of the same
simulation provide qualitative insight into the impact of AGN on the observed star-formation rate versus stellar mass plane
(Fig.~4). In the model, it can be seen that AGN are responsible for
creating a shallower slope at the highest stellar masses as well as
reducing the overall number of massive galaxies\cite{Schaye15,Crain15}. The builders
of the Horizon-AGN hydrodynamical cosmological simulation recently performed a similar
test by running the simulation with and without AGN feedback and came to the
same conclusion: the effect of AGN is to significantly reduce the
star formation rates of massive galaxies with the magnitude of suppression increasing with
stellar mass\cite{Beckmann17}. Therefore, it appears that the observational signature of AGN suppressing star formation may be imprinted on the reduced
average star formation rates per unit stellar mass for the most massive galaxies
(Fig.~4) and {\em not} on reduced average star formation rates
for the most instantaneously luminous AGN (Fig.~3). 

The results described above, and other recent work, highlight that investigating the star formation properties for populations of massive galaxies, not just
AGN-host galaxies, at multiple cosmic epochs is a critical test for different AGN feedback
prescriptions\cite{Thacker14,Bongiorno16,Bluck16,Terrazas16}.

\section*{Conclusions}
Some of the key conclusions brought up in this review are:

(1) Local mechanically-dominated AGN are energetically capable of
  regulating gas cooling on large scales via radio jets in the most massive
  haloes and consequently regulating star formation inside their host
  galaxies. However, it is uncertain what ``quenched'' the high levels of star-formation that previously
  occurred in these galaxies and what role these AGN play at early
  cosmic epochs ($z\gtrsim0.5$) and in less dense environments.  

(2) Radiative AGN are observed to be driving outflows in multiple phases of
  gas. For many galaxies, measurements of energy and mass outflow rates
  have implied that star formation could be suppressed by the removal
  of star-forming material. However, the long-term
  impact of these events is unclear. In a few cases AGN-driven jets
  are also observed to be triggering local episodes of star
  formation. 

(3) The suppression or regulation of star-formation by an AGN does
  not need to be the result of the complete evacuation of gas from a
  galaxy. Observations of turbulence, shocks and heating by AGN jets and outflows
 suggest that they are able to reduce the efficiency of converting the
 available gas supply into stars without the need to remove it.

(4) The most massive galaxies ($M_{\rm stellar}\gtrsim10^{10}$\,M$_{\odot}$)
  have low star formation rates per unit stellar mass across
  multiple cosmic epochs. Although not conclusive,
  this could be due to star formation suppression by the cumulative
  effect of AGN episodes.

(5) The timescales of various feeding and feedback
processes remain uncertain. For example, AGN may no longer be visible or
luminous when the impact that they have had becomes observable.  Consequently, great care must be taken when using empirical
results to draw conclusions on ``smoking gun''  evidence for or
against the impact of AGN upon star formation.  Whilst we may observe
the ``smoke'' (e.g., outflows and/or reduced star formation rates) the ``gun'' (i.e., the AGN) may no longer be visible. 

\section*{Future prospects}
Further work combining {\em specific} theoretical predictions with observations is
required to make significant progress in understanding the long term impact of AGN on their
host galaxies. Hydrodynamical cosmological models provide the
means to make predictions on the star-formation properties and their evolution of statistical
samples of galaxies using a variety of feedback models. In parallel to this, high-resolution
simulations can indicate what the observational signatures are for various
mechanisms of how AGN could transfer energy and
momentum into the gas in individual galaxies. 

From observations, over the next five to ten years we can expect to see considerable progress
in the number of high-quality measurements to test these models. For example, the upgrade of (sub)-millimetre interferometers such as
ALMA and NOEMA will produce sensitive, high resolution observations of dust
emission and molecular gas in an increasing number of sources across
multiple cosmic epochs. Such observations will significantly reduce
the uncertainties on derived quantities such as star
formation rates and mass outflow rates. Forthcoming facilities
such as {\em JWST} (due to be launched in 2018) and 30m-class
telescopes (expected first light in the early 2020s) will enable us
measure gas inflows, outflows and host galaxy properties (such as
stellar masses and star-formation histories), to unprecedented precision for large samples of
extremely distant galaxies ($z\gg1$). Furthermore, the data from {\em
  eROSITA} (due to be launched in 2018) will yield X-ray
identification of millions of AGN, which could provide a key role in testing model predictions
on large, statistical samples of AGN host galaxies. 

\section*{Acknowledgements} 
I acknowledge the referees for their constructive input and the Science and Technology Facilities Council through
grant code ST/L00075X/1. Thanks go to Dave Alexander, David Rosario,
Stuart McAlpine and James Mullaney for stimulating discussion. Thanks also go to the EAGLE
consortium for making the data from their simulations public.

\phantomsection
\bibliographystyle{naturemag}


\end{document}